\documentclass[final,5p,times,twocolumn]{elsarticle}

\journal{New Astronomy}

\biboptions{authoryear}

\usepackage{color}

\usepackage{natbib}

\usepackage{graphicx}	
\usepackage{amsmath}	
\usepackage{amssymb}	
\usepackage{gensymb}

\begin{document}

\begin{frontmatter}

\title{Variable stars in Palomar 13; an evaporating globular cluster.}

\author[IA]{M. A. Yepez\corref{mycorrespondingauthor}}
\cortext[mycorrespondingauthor]{Corresponding author}
\ead{myepez@astro.unam.mx}

\author[IA]{A. Arellano Ferro}
\ead{armando@astro.unam.mx}

\author[DAUG]{K.P. Sch{\"o}eder}
\ead{kps@astro.ugto.mx}

\author[IIA]{S. Muneer}
\ead{muneers@iiap.res.in}

\author[IIA]{Sunetra Giridhar}
\ead{sunetragiridhar@gmail.com}

\author[IA]{Christine Allen}
\ead{chris@astro.unam.mx}

\address[IA]{Instituto de Astronom\'ia, Universidad Nacional Aut\'onoma de M\'exico, Ciudad Universitaria, 04510, M\'exico.}

\address[DAUG]{Departamento de Astronom\'ia, Universidad de Guanajuato, M\'exico}

\address[IIA]{Indian Institute of Astrophysics, Bangalore, India.}

\begin{abstract}
We present new CCD $VI$ photometry of the distant globular cluster Pal 13. Fourier decomposition of the light curves of the three cluster member RRab stars lead to estimations of [Fe/H]=-1.65, and a distance of 23.67$\pm$0.57 kpc. Light and colour near minimum phases for RRab stars leads to an estimate of $E(B-V)$=0.104 $\pm$ 0.001. A $V/(V-I)$ colour-magnitude diagram, built exclusively with likely star members, shows consistency with the above parameters and an age of 12 Gyrs. A search of variable stars in the field of view of our images revealed the variability of a red giant cluster member and of three probably non-member stars; two RRab stars and one W Virginis star or CW. The GAIA proper motions of member stars in Pal 13 show a significant scatter, consistent with the scenario of the cluster being tidally stripped.
\end{abstract}

\begin{keyword}
globular clusters: individual (Pal 13) -- stars:variables:RR Lyrae
\end{keyword}

\end{frontmatter}

\section{Introduction}

Palomar 13 is among a group of outer Galactic halo globular clusters (GC) that stand out for their low luminosity/radius ratio. Other members of the group are AM-4, Koposov 1,2, Whiting 1, Pal 1 and Seg 3 \citep{Fadely2011}.

Its low luminosity and rather large size, $M_v=-3.7$ and tidal radius $R_t= 23.3$ pc respectively \citep{Mackey2005}, place it half way between Milky Way satellite  ultra-faint galaxies and  Galactic GC \citep{Martin2008,Bradford2011} and the observed large radial velocity dispersion was interpreted as due to the tidal heating during perigalacticon passage or to the presence of dark matter \citep{Cote2002}.
An alternative explanation is the presence of a large fraction of unresolved binary stars that increase the single-epoch radial velocity dispersion  \citep{Blecha2004}.

The fraction of binary stars is larger in clusters of low-luminosity and they can have a population in excess $\sim$ 40 percent in the cluster core \citep{Milone2016}. According to \cite{Clark2004}, Pal 13 exhibits a high binary fraction of about 30\%$\pm$4\%.

However, when the velocity dispersion is corrected for the presence of unresolved binary stars, and the radial spread of metallicities is compared with that in ultra-faint galaxies, it is suggested that the cluster dynamics is consistent with the stellar mass and no dark matter or extreme tidal heating are required in this low-mass cluster \citep{Bradford2011}.
It has been suggested that Pal 13 is approaching its apogalacticon (farthest from Galactic center), and that tidal debris due to stripping  throughout its orbit are compressed back to the cluster as it decelerates \citep{Kupper2011}. The effects of orbital eccentricity and inclination on globular cluster tidal radii, escaping stars, and tidal compression at perigalacticon has also been discussed by \citet{Webb2014a,Webb2014b}. 

Recently available GAIA proper motions \citep{Gaia2018} should shed some light on this scenario. 

The $V/(V-I)$ colour-magnitude diagram (CMD) of Pal~13 displays a very scantily populated horizontal branch (HB) with no blue tail at all. The instability strip is occupied by only four RR Lyrae stars known at present. It seems likely that given the structure of the HB, no other RR Lyrae star is present in the cluster. The CMD shows however a significant population of blue stragglers, a number of bright stars in the upper instability strip above the HB, and some bright red giant stars, among which a thorough search might be rewarded with the detection of a few more variables. Since the discovery of the four RR Lyrae (V1-V4) by \cite{Rosino1957}, a very few time-series studies have been dedicated to the cluster; e.g. \cite{Ciatti1965}, \cite{Ortolani1985}, \cite{Siegel2001}, the later already in the CCD era. However, these studies are based on a handful of plates or images taken in a short-time span. Surprisingly, the light curves of the four known RR Lyrae have not been studied in detail. We believe that a dedicated effort to study the morphology of the light curves and their use as physical parameters indicators, as well as a search for possible new variables, is in order. Despite the cluster being a distant and faint one, the common variable stars regions in the CMD, such as the Blue Stragglers, the RGB and upper instability strip, are within the reach of deep CCD photometry. In the present work we report the findings from the extended time-series photometry of this cluster and the exploration of the proper motions of stars in the field of Pal 13.

\begin{table}[t]
\footnotesize
\caption{The distribution of observations of Pal 13.$^{*}$}
\centering
\begin{tabular}{@{}lccccc}
\hline
Date  &  $N_{V}$ & $t_{V}$ (s) & $N_{I}$ &$t_{I}$ (s)&Avg seeing (") \\
\hline
20111004 & 16 &   600   & 18 &   200   & 2.0\\
20111005 & 20 &   600   & 20 &   200   & 2.4\\
20111006 &  7 &   600   &  8 &   200   & 2.5\\
20111007 &  1 &   600   &  2 &   200   & 1.7\\
20111102 & 16 & 400-600 & 16 & 125-200 & 1.5\\
20111103 & 14 & 380-450 & 14 & 125-200 & 1.4\\
20111104 &  9 & 425-500 & 10 & 120-200 & 1.6\\
20111105 &  0 &   ---   &  3 &  60-120 & 1.6\\
20111215 & 12 & 500-600 & 15 & 125-200 & 2.8\\
20111216 &  3 &   600   &  4 &   200   & 1.9\\
20131024 & 19 & 300-400 & 19 & 120-150 & 1.7\\
20131025 &  8 &   480   & 10 &   240   & 2.0\\
20141015 &  2 &   700   &  4 &   250   & 1.8\\
20141016 &  4 &   500   &  6 &   250   & 1.8\\
20141017 &  2 &   500   &  2 &   250   & 3.4\\
20180712 & 29 &    60   & -- &    --   & 1.7\\
20180713 & 24 &    60   & -- &    --   & 1.5\\
20180714 & 32 &    60   & -- &    --   & 1.4\\
20180715 &  4 &    60   & -- &    --   & 1.6\\
20180724 & 20 &    60   & -- &    --   & 1.7\\
20180725 & 23 &    60   & -- &    --   & 1.4\\
20180811 & 56 &    60   & -- &    --   & 1.9\\
20180812 & 72 &    60   & -- &    --   & 1.5\\
20180813 & 47 &    60   & -- &    --   & 1.8\\
20180814 & 72 &    60   & -- &    --   & 1.5\\
20180815 & 64 &    60   & -- &    --   & 1.6\\
\hline
Total:   & 576 &        & 151 &           &\\
\hline
\end{tabular}
\center{*: Columns $N_{V}$ and $N_{I}$ give the number of images taken with the $V$ and $I$ filters respectively. Columns $t_{V}$ and $t_{I}$ contain the exposure time. The average seeing is listed in the last column.}
\label{tab:observations}
\end{table}

\section{Observations and Reductions}
\label{sec:Observations}

\begin{figure*} 
\includegraphics[width=18.0cm]{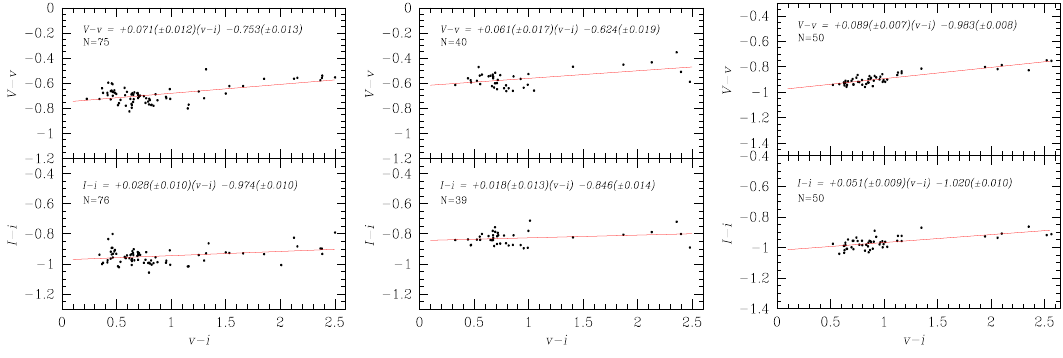}
\caption{Colour dependence of the standard minus instrumental systems and the
transformation equations for the three settings of our images, each based on N local
standards \citep{Stetson2000}.}
    \label{transV}
\end{figure*}

\subsection{Observations}

The observations were performed from two sites; 1) on 15 nights between October 4,
2011 and October 17, 2014 with the 2.0 m telescope at the Indian Astronomical
Observatory (IAO), Hanle, India. A total of 133 and 151 images were obtained  in the
Johnson-Kron-Cousins $V$ and $I$ filters, respectively. The detector was a SITe ST-002
thinned backside illuminated CCD of
 2048$\times$2048 pixels with a scale of 0.296 arcsec/pix, translating to a field of
view (FoV) of approximately 10.1$\times$10.1~arcmin$^2$. 
2) on 11 nights between July 12, 2018 and August 15, 2018 with the 0.84 m telescope at
the Observatorio Astron\'omico Nacional on the Sierra San Pedro M\'artir (OAN-SPM),
Baja California, M\'exico. A total of 443 images were obtained  in the Johnson $V$
filter. The detector was a Spectral Instruments CCD of 1024$\times$1024 pixels with a
scale of 0.444 arcsec/pix, translating to a field of view (FoV) of approximately
7.57$\times$7.57~arcmin$^2$. 

The log of observations is given in Table \ref{tab:observations} where the dates,
number of frames, exposure times and average nightly seeing are recorded. 

\subsection{Difference Image Analysis}
We employed the technique of difference image analysis (DIA) (\citealt{Alard2000},
\citealt{Bramich2005}) to carry out high-precision photometry for all of the point
sources in the images of Pal 13 and we used the {\tt DanDIA}\footnote{DanDIA is built
from the DanIDL library of IDL routines
available at http://www.danidl.co.uk} pipeline for the data reduction process
\citep{Bramich2008,Bramich2013}. For the data sets of each observatory a reference
image for the $V$ filter and another for the $I$ filter were constructed by stacking
the best-quality images in each collection; then sequences of difference images in
each filter were built by subtracting the respective convolved reference image from
the rest of the collection. Differential fluxes for each star detected in the
reference image were then measured on each difference image. 
Light curves for each star were constructed by calculating the total 
flux $f_{\mbox{\scriptsize tot}}(t)$ in ADU/s at each epoch $t$ from:
\begin{equation}
f_{\mbox{\scriptsize tot}}(t) = f_{\mbox{\scriptsize ref}} +
\frac{f_{\mbox{\scriptsize diff}}(t)}{p(t)},
\label{eqn:totflux}
\end{equation}
where $f_{\mbox{\scriptsize ref}}$ is the reference flux (ADU/s),
$f_{\mbox{\scriptsize diff}}(t)$ is the differential flux (ADU/s) and
$p(t)$ is the photometric scale factor (the integral of the kernel solution).
Conversion to instrumental magnitudes was achieved using:
\begin{equation}
m_{\mbox{\scriptsize ins}}(t) = 25.0 - 2.5 \log \left[ f_{\mbox{\scriptsize tot}}(t)
\right],
\label{eqn:mag}
\end{equation}
where $m_{\mbox{\scriptsize ins}}(t)$ is the instrumental magnitude of the star 
at time $t$. 
  The above procedure has been described in detail in \cite{Bramich2011}.

\begin{table}
\footnotesize
\caption{Time-series $V$ and $I$ photometry for all the variables in the field of view
of Pal 13. (Full table is available in electronic format).}
\centering
\begin{tabular}{ccccc}
\hline
Variable &Filter & HJD & $M_{\mbox{\scriptsize std}}$ &
$\sigma_{m}$  \\
Star ID  &        & (d) & (mag) &(mag)\\
\hline
V1& V & 2455839.17019&17.706& 0.003\\
V1& V & 2455987.37903&17.664& 0.003\\
\vdots   & \vdots & \vdots  & \vdots & \vdots  \\
V1& I & 2458345.99782&17.299& 0.013\\
V1& I & 2455839.17700&17.091&0.006 \\
\vdots   & \vdots & \vdots  & \vdots & \vdots  \\
V2& V & 2455839.17019&17.899& 0.004\\
V2& V & 2455839.20103&17.949& 0.004\\
\vdots   & \vdots & \vdots  & \vdots & \vdots \\
V2& I & 2455839.16202&17.181& 0.012\\
V2& I & 2455839.17700&17.171& 0.007\\
\vdots   & \vdots & \vdots  & \vdots & \vdots \\
\hline
\end{tabular}
\label{tab:vi_phot}
\end{table}

\begin{figure*}
\includegraphics[width=17.0cm]{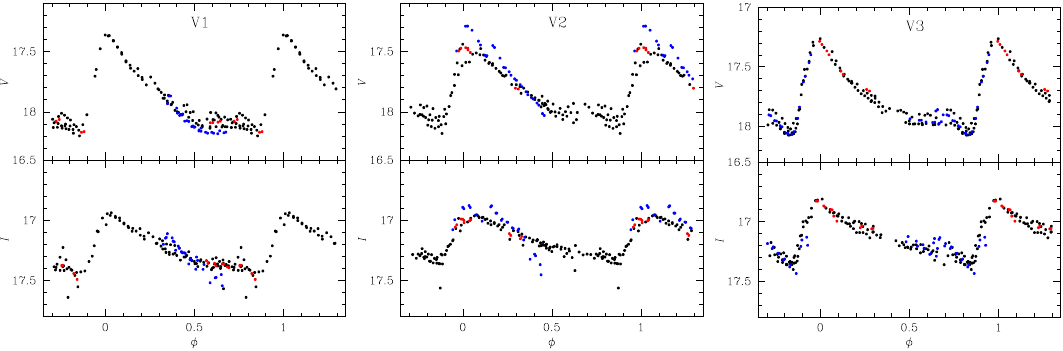}
\caption{Light curves of three RRab stars in Pal 13 from the Hanle data phased with
seasonal periods calculated exclusively from this data-set. The colours distinguish
data from each setting and enhance the amplitude modulations.}
\label{curvasPal13}
\end{figure*}

\begin{figure}
\includegraphics[width=8.5cm]{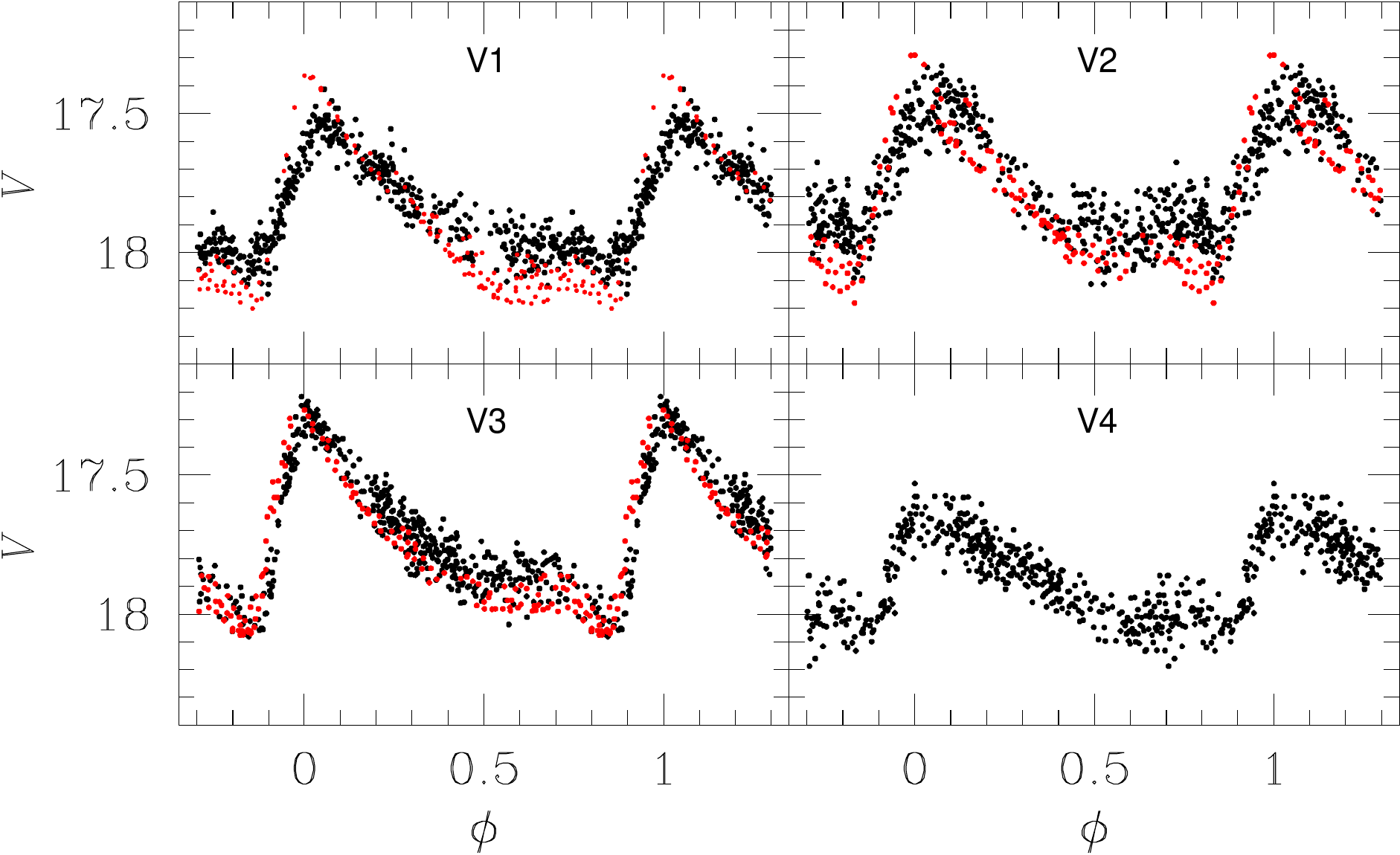}
\caption{$V$ light curves of the 4 RRab stars in Pal 13 phased with the refined
periods given in Table \ref{variables}. Black and red symbols refer to data from SPM
and Hanle respectively. V4 is not in the FoV of Hanle images.}
\label{lc_HANSPM2}
\end{figure}

\subsection{Photometric Calibrations}

\subsubsection{Relative calibration}
\label{sec:rel}

To correct from possible systematic errors and avoid mistaking them as  bona fide
variability in light curves, we apply the methodology developed in \cite{Bramich2012}
to solve for the magnitude offsets $Z_{k}$ that should be applied to each photometric
measurement from the image $k$. In terms of DIA, this translates into a correction (to
first order) for the systematic error introduced into the photometry from an image due
to an error in the fitted value of the photometric scale factor $p$
\citep{Bramich2015}. We found that, in the present case, the systematic error
corrections were negligible even for the brighter stars.

\subsubsection{Absolute calibration}
\label{absolute}

Standard stars in the field of Pal 13 are included in the online collection of
\cite{Stetson2000}
\footnote{http://www3.cadc-ccda.hia-iha.nrc-cnrc.gc.ca/community/STETSON/standards}
and we used them to transform instrumental $vi$ magnitudes into the standard \emph{VI}
system.

For the observations of Hanle, three different pointings were used. Since the
collection of standard stars changes for each setting, we performed an independent
transformation for each case. In all cases the standard minus the instrumental
magnitudes show mild dependencies on the colour, as can be seen in Fig.\ref{transV}.
The transformation equations between the instrumental and the standard magnitudes are
given in each panel of the figure. Similar transformations were calculated for the
observations of SPM and for briefness they are not explicitly included in this paper.

All of our \emph{VI} photometry for the known variable stars in the FoV of Pal 13 is
provided in Table \ref{tab:vi_phot}. A small portion of this table is given in the
printed version of this paper and the full table is available in electronic form. 

\begin{table*}
\footnotesize
\caption{General data for the variable stars in the FoV of Pal 13. Variables with
Blazhko modulations are labeled '$Bl$'. Amplitudes for Blazhko variables correspond to
the maximum observed.}
\label{variables}
\centering
\begin{tabular}{clcccccccc}
\hline
Variable & Variable & $<V>$   & $<I>$   & $A_V$       & $A_I$   & $P$ (days)    &
 HJD$_{\rm max}$     &  RA          & Dec.         \\
Star ID  & Type$^{a}$& (mag)   & (mag)   & (mag)       & (mag)   & this work     
&  (d +245~0000.)    &  (J2000.0)   & (J2000.0)  \\
\hline
V1 & RRab $Bl$ & 17.886 & 17.244 & 0.80 & 0.51 & 0.538192 & 5839.2010 & 23:06:41.53 &
+12:46:58.94 \\
V2 & RRab $Bl$ & 17.819 & 17.164 & 0.79 & 0.51 & 0.597080 & 6591.2802 & 23:06:44.47 &
+12:46:16.12 \\
V3 & RRab      & 17.766 & 17.121 & 0.83 & 0.56 & 0.578190 & 5839.2998 & 23:06:43.14 &
+12:46:47.80 \\
V4 & RRab $Bl$ & 17.886 &    -   & 0.52 &   -  & 0.575130 & 8342.8849 & 23:06:48.83 &
+12:41:22.30 \\
V5$^{1}$ & SR        & 17.029 & 15.930 & 0.4  &   -  & 7.26 & 5842.1160 & 23:06:37.28
& +12:49:27.49 \\
$Var1^{1,*}$ & RRab    & 21.123 & 19.720 & 1.6  &   -  & 0.508228 & 5870.0997 &
23:06:30.25 & +12:49:40.16 \\
$Var2^{1,*}$ & RRab      & 17.993 & 16.731 & 1.1  &   -  & 0.515811 & - & 23:06:33.32
& +12:42:41.61 \\
$Var3^{1,*}$ & CW      & 17.452 & 16.619 & 0.3  &   -  & 7.57 & 5841.2595 &
23:06:41.44 & +12:49:28.71 \\ 
\hline
\end{tabular}
\center{1: Newly found in this paper.\\
*: These stars are very likely not cluster members.}
\end{table*}

\section{The RR Lyrae stars in Pal 13}

Pal 13 being a very inconspicuous globular cluster, published light curves of the RR
Lyrae V1-V4 are very rare. To the best of our knowledge, only blue light curves were
published by \cite{Ciatti1965}. Unfortunately the data used in their work are not
available for systematic comparison. Thus, we believe this is the first detailed
analysis of the light curves of the 4 RR Lyrae known in the field of the cluster. A
discussion of the cluster physical parameters from its RR Lyrae stars and the stellar
membership in the cluster follows in the remainder of this work.

\subsection{Periods}
\label{sec:Periods}

The periods of V1, V2 and V3, estimated using exclusively the data from the Hanle
data-set properly phase the light curves as shown in Fig. \ref{curvasPal13}. These
periods however, if applied to the SPM data-set produce light curves phase-shifted by
a few tenths of a phase suggesting that the periods need refinement. The combination
of the two data sets provide the refined periods listed in Table \ref{variables},
producing the phased light curves in Fig. \ref{lc_HANSPM2}. The panels show the good
correspondence of the Hanle (red) and SPM (black) light curves. For V4 we have no
Hanle data since it was not included in the FoV.

\subsection{Reddening}
\label{sec:redd} 

Before proceeding to the physical parameter estimations, it is necessary to decide the
reddening of the cluster. We have taken advantage of the result from \cite{Sturch1966}
that RRab stars have nearly the same intrinsic colour $(B-V)_0$ at minimum light, and
of the calibration made by \cite{Guldenschuh2005} of $(V-I)_{0;min}$ = 0.58$\pm$0.02.
Applying these results to the four RRab stars in Pal 13 we found an average value of
$E(B-V)=$0.104$\pm$0.001, in excellent agreement with the the values from the dust
maps and the calibrations of \cite{Schlegel1998} and \cite{Schlafly2011} of
0.114$\pm$0.004 and 0.098$\pm$0.003, respectively. We shall adopt $E(B-V)=$0.10, or
$E(V-I)=1.259 E(B-V)$ \citep{Schlegel1998}, in the rest of this paper.

\subsection{Light curve Fourier decomposition}
\label{sec:RRLstars}

With the aim of estimating key physical parameters, a Fourier decomposition of the
light curves of the RR Lyrae stars was performed.

To this end, the standard procedure is to represent the $V$ light curves by the
Fourier series of harmonics:

\begin{equation}
m(t) = A_0 ~+~ \sum_{k=1}^{N}{A_k ~\cos~( {2\pi \over P}~k~(t-E_0) ~+~ \phi_k ) },
\label{eq_foufit}
\end{equation}

\noindent
where $m(t)$ is the magnitude at time $t$, $P$ is the period and $E_0$ the epoch,
generally a time of maximum light. A
linear minimization routine is used to derive the amplitudes $A_k$ and phases $\phi_k$
of each harmonic, from which the Fourier parameters $\phi_{ij} = j\phi_{i} -
i\phi_{j}$ and $R_{ij} = A_{i}/A_{j}$ are calculated. The resulting mean magnitudes
$A_0$, and the Fourier coefficients  for the four RRab stars are listed in Table
~\ref{fourier_coeffs}. While in principle we should not include stars with evident
amplitude-phase modulations. Since the RR Lyrae sample is limited, we made an effort
to include the Blazhko variables V1, V2 and V4 by fitting their data at largest
amplitude. It is well known that if Blazhko and stable stars are mixed in the
calculation of physical parameters the uncertainties will increase. Particularly the
value of [Fe/H], which if determined 
from phases of low amplitudes will be overestimated \citep{Cacciari2005}. According to
\cite{Szeidl1976}, Blazhko variables at largest amplitudes are comparable to stable
stars. Hence, we decomposed these Blazhko RR Lyrae light curves exclusively from data
at maximum amplitude.

\begin{table*}
\footnotesize
\caption{Fourier coefficients $A_{k}$ for $k=0,1,2,3,4$, and phases $\phi_{21}$,
$\phi_{31}$ and $\phi_{41}$, for RRab stars. The numbers in parentheses indicate the
uncertainty on the last decimal place. Also listed is the deviation parameter
$D_{\mbox{\scriptsize m}}$ (see Section~\ref{sec:RRLstars}).}
\centering                   
\begin{tabular}{lllllllllr}
\hline
Variable ID & $A_{0}$ & $A_{1}$ & $A_{2}$ & $A_{3}$ & $A_{4}$ & $\phi_{21}$ &
$\phi_{31}$ & $\phi_{41}$ &  $D_{\mbox{\scriptsize m}}$ \\
   & ($V$ mag)  & ($V$ mag)  &  ($V$ mag) & ($V$ mag)& ($V$ mag) & & & & \\
\hline
V1 & 17.886(4) & 0.310(6) & 0.114(6) & 0.100(6) & 0.058(6) & 3.882( 62) & 8.104( 80) &
5.974(122) & 2.1\\
V2 & 17.819(5) & 0.279(7) & 0.118(7) & 0.091(8) & 0.049(7) & 3.747( 81) & 8.191(108) &
6.175(178) & 2.9\\
V3 & 17.766(4) & 0.267(6) & 0.131(5) & 0.101(5) & 0.062(6) & 3.765( 61) & 7.890( 86) &
5.875(123) & 2.7\\
V4 & 17.886(3) & 0.173(4) & 0.051(4) & 0.049(5) & 0.027(4) & 3.990(103) & 8.169(120) &
6.608(193) & 1.6\\
\hline
\end{tabular}
\label{fourier_coeffs}
\end{table*}

These Fourier parameters and the semi-empirical calibrations of \cite{Jurcsik1996},
for RRab stars, were used to obtain  [Fe/H]$_{\rm ZW}$ on the \cite{Zinn1984}
metallicity scale that can be transformed to the UVES scale using the equation
[Fe/H]$_{\rm UVES}$=$-0.413$ +0.130~[Fe/H]$_{\rm ZW}-0.356$~[Fe/H]$_{\rm ZW}^2$
\citep{Carretta2009}. The absolute magnitude $M_V$ can be derived from the
calibrations of \cite{Kovacs2001} for RRab stars. $T_{\rm eff}$ was estimated using
the calibration of \cite{Jurcsik1998}. For brevity we do not explicitly present here
the above mentioned calibrations; however, the corresponding equations, and most
importantly their zero points, have been discussed in detail in previous papers
\citep{Arellano2011,Arellano2013}, where the corresponding equations towards the
calculation of $T_{\rm eff}$, log$(L/{\rm L_{\odot}})$, $M/{\rm M_{\odot}}$ and
$R/{\rm R_{\odot}}$ can be found. We refer the interested reader to those papers.

\noindent 
The [Fe/H] calibration for RRab stars of \cite{Jurcsik1996} is applicable to stars
with a {\it deviation parameter} $D_m$, defined by \cite{Jurcsik1996} and
\cite{Kovacs1998}, not exceeding an upper limit. These authors suggest $D_m \leq 3.0$.
The $D_m$ is listed in column 10 of Table~\ref{fourier_coeffs}. The criterion is
fulfilled by all four stars.

\begin{table*}
\footnotesize
\caption[] {\small Physical parameters for the RRab stars. The numbers in parentheses
indicate the uncertainty on the last decimal place.}
\centering
\label{fisicos}
\hspace{0.01cm}
 \begin{tabular}{ccccccccc}
\hline 
Star&[Fe/H]$_{\rm ZW}$  &[Fe/H]$_{\rm UVES}$ &$M_V$ & log~$T_{\rm eff}$  &
log~$(L/{\rm L_{\odot}})$ & $D$ (kpc) & $M/{\rm M_{\odot}}$&$R/{\rm R_{\odot}}$\\
\hline
V1 & $-1.50(8)$ & $-1.41(8)$ & 0.638(9) & 3.812(15) & 1.645(3)& 24.27(10) & 0.67(12) &
5.31(2) \\
V2 & $-1.64(10)$& $-1.58(12)$& 0.582(10)& 3.804(21) & 1.667(4)& 24.15(12) & 0.67(17) &
5.65(3) \\
V3 & $-1.85(8)$ & $-1.88(11)$& 0.629(8) & 3.801(15) & 1.648(3)& 23.06(9) & 0.69(13) &
5.59(2) \\
V4 & $-1.58(11)$& $-1.51(13)$& 0.701(6) & 3.800(23) & 1.620(2)& 23.58(7) & 0.65(18) &
5.45(2) \\
\hline
Weighted Mean & $-1.65(4)$ & $-1.58(5)$ & 0.654(4) & 3.805(9) & 1.639(2) & 23.67(4) &
0.67(7) & 5.48(1)\\
$\sigma$&  $\pm$0.15  &  $\pm$0.20  &$\pm$0.052&$\pm$0.005&$\pm$0.019 & $\pm$0.57
&$\pm$0.02&$\pm$0.15 \\
\hline 
\end{tabular}
\center
\end{table*}

The individual physical parameters for the four RRab stars and the inverse-variance
weighted means are reported in Table~\ref{fisicos}.

The average [Fe/H]$_{\rm ZW}$=$-1.65\pm0.15$ is in agreement with the average
spectroscopic determination of [Fe/H]=$-1.66\pm 0.1$ found by \cite{Bradford2011} on
16 member stars.

The weighted mean $M_V$ values for the RRab stars are 0.621$\pm$0.031 mag and will be
used in section \ref{sec:distance} to estimate the mean distance to the parent
cluster.

\section{The Oosterhoff type of Pal 13}
\label{OoType}

The value of [Fe/H]=-1.88 for Pal 13 adopted by \cite{Harris1996} in his data
compilation of globular clusters immediately suggests an Oosterhoff type II for the
cluster (OoII), as no OoI cluster is known with such a low metallicity. Nevertheless,
the average period of its RRab stars V1-V4 is 0.572 days, i.e. the expected value for
an OoI cluster.

\begin{figure} 
\includegraphics[width=8.cm,height=13.cm]{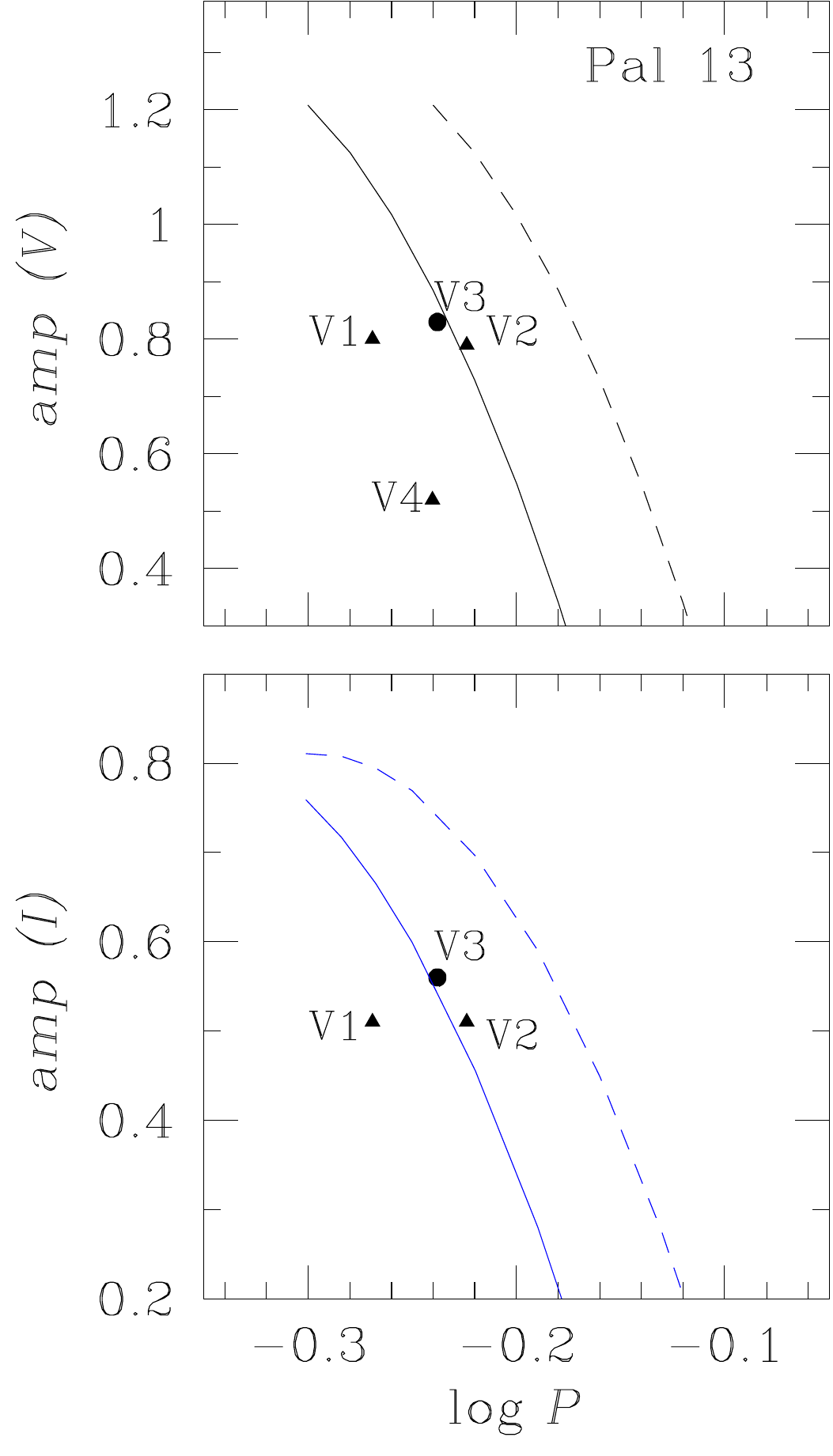}
\caption{The log $P$ vs. amplitude plane for the $V$- and $I$-bands of the four RRab
stars in Pal 13. Triangles represent stars with hints of Blazhko modulations. In the
top panel the continuous and segmented lines are the loci found by \cite{Cacciari2005}
for unevolved and evolved stars, respectively, in the OoI cluster M3. In the bottom
panel the blue loci are from \cite{Kunder2013a} for the OoI cluster NGC~2808.}
\label{Bailey}
\end{figure}

The Bailey, or period-amplitude diagram, of Pal 13 for the $V$ and $I$ filters is
shown in Fig. \ref{Bailey}. In this diagram the continuous lines represent the loci
derived for the OoI clusters M3 by \cite{Cacciari2005} in $V$-band, and NGC 2808 by
\cite{Kunder2013a} in $I$-band. dashed curves correspond to the loci that should
occupy RR Lyrae stars that are evolved past the ZAHB, typical of OoII systems. Despite
the low number of RRab stars in Pal 13 and the fact that some are affected by Blazhko
modulations, it is clear that their distributions corresponds to that of an OoI
cluster.

Our estimation of [Fe/H]$_{\rm ZW}$=-1.65$\pm $0.15 places Pal 13 among the most metal
week of the OoI clusters, competing with NGC 4147 \citep{Arellano2018_N4147}.

\section{Distance to Pal 13 from the RR Lyrae stars}
\label{sec:distance}
 
We calculate the distance to Pal 13 using two independent approaches based on the RR
Lyrae stars. The first approach involves the calculation of $M_V$ via the Fourier
light curve decomposition of the RRab stars. We averaged the distances only for V1-V4,
to find 23.67 $\pm$ 0.57 kpc. The membership of V4 to the cluster has been doubted
given its large distance from the cluster center. We note however, that the distance
to V4 itself, 23.72 kpc, is not at all different from the other three RRab stars. It
should be noted too that V4 at about 5' from the cluster center, is well within the
King tidal radius, which according to \cite{Bradford2011} is at 13.9'. We shall offer
further arguments in favour of V4 cluster membership later in this paper.

The second approach is from the $I$-band RR Lyrae P-L relation derived by
\cite{Catelan2004}: 

\begin{equation}
M_I=0.471-1.372 \log P + 0.205 \log Z
\label{eqn:PL_RRI}
\end{equation}
with $\log Z =\rm [M/H] - 1.765$ and $\rm [M/H]= [Fe/H] - \log(0.638 f+0.362)$ and
$\rm \log f =[\alpha/Fe]$ \citep{Salaris1993}.

We do not have $I$-band data for V4. The average result for V1-V3 is 24.34 $\pm$ 0.50
kpc.

The above independent estimations of the cluster distance are, within their
uncertainties, in good agreement.

\section{The Colour-Magnitude Diagram of Pal 13}
\label{sec:CMD}

\begin{figure} 
\includegraphics[width=8.5cm,height=8.5cm]{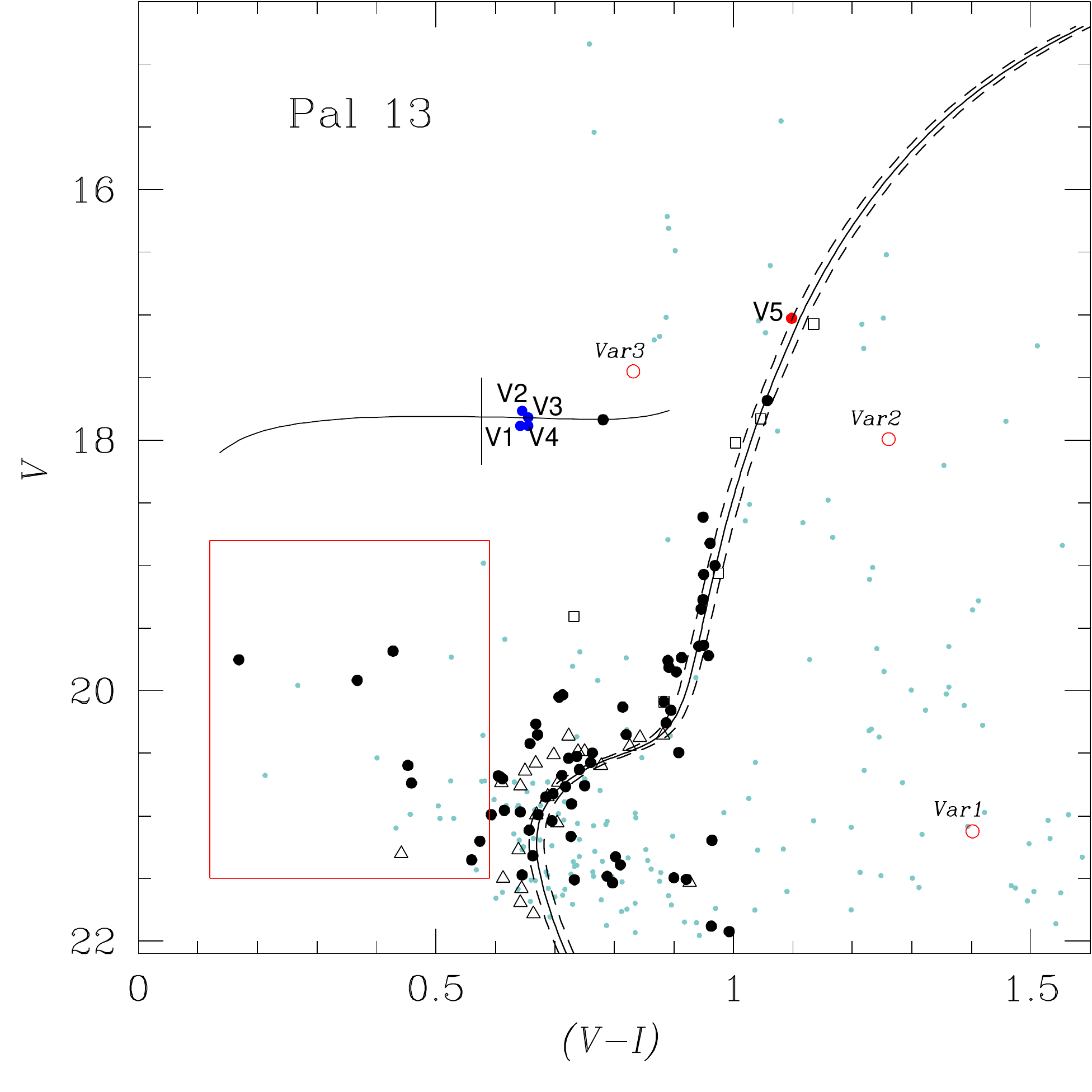}
\caption{Colour-Magnitude diagram of Pal13 based on our data from Hanle. The black
dots represent the member stars according to Siegel et al. (2012). Empty squares are
members according to their radial velocity analysis (Cot\'e et al. 2012), and the
empty triangles are members according to the most recent proper motion analysis in the
SIMBAD data base.  Isochrones of 12 Gyrs, and ZAHB are from \cite{VandenBerg2014} for
[Fe/H]=$-1.65$, $Y=0.25$ and [$\alpha$/Fe]=0.4 shifted to the reddenings, from left to
right, of $E(V-I)$=0.10, 0.11 and 0.12  and to a distance of 23.67 kpc. The red
rectangle is the Blue Straggler region where our search for SX Phe stars was carried
out. See $\S$ \ref{sec:CMD} for a discussion.}
    \label{CMD}
\end{figure}

Pal 13 is a faint and sparse cluster, hence the field is populated by a large number
of non-member stars. A proper motion photographic analysis by \cite{Siegel2001}
identified a number of reliable cluster members. We built a $V/(V-I)$ Colour-Magnitude
Diagram (CMD) in Fig. \ref{CMD} where the black symbols are member stars while the
light blue are not. We complemented the list of member stars by a few identified by
\cite{Cote2002} from their radial velocities and from the list of members furnished
from the proper motions available at the SIMBAD Pal 13 data base; open squares and
open triangles in Fig. \ref{CMD}, respectively. A total of 102 members with $VI$
photometry in our study have been identified.

It is obvious that the cluster lacks of a significant red giant branch (RGB), and the
HB is only populated by five stars including the four RR Lyrae, a fact that was
noticed early in the  study of \cite{Ciatti1965}. The black vertical line in the HB of
Fig. \ref{CMD}, represents the border between the first overtone and fundamental
pulsating RR Lyrae, or the red edge of the first overtone (RFO) instability strip
\citep{Arellano2016}. 
We do not have $I$-band data for V4 as it is not in the field of Hanle and no $I$
observations were done in SPM. We have estimated its $V-I$ colour from a few images
taken at the SWOPE telescope in Las Campanas and kindly made available to us by Dr.
Nidia Morrell. V4 is found next to V1-V3 in the CMD, as expected given that it has a
very similar temperature (see Table \ref{fisicos}).
Thus, V1-V4 fall on the fundamental region and not on the "either-or" region, to the
blue of the RFO. Pal 13 is one more example of Oo I type clusters without RRab stars
in the "either-or" region, a property shared with all Oo II and with some but not all
OoI clusters; the exceptions being for example with NGC 3201, NGC 5904 and NGC 6934,
see \cite{Arellano2018_N6171} for a discussion (their figure 8).

The age of Pal 13 has been approximately estimated as 12 Gyrs,
\citep{Borissova1997,Bradford2011}, thus we overlaid an isochrone for this age and a
zero age horizontal branch (ZAHB) from the model collection of \cite{VandenBerg2014}
for [Fe/H]=$-1.65$, $Y=0.25$ and [$\alpha$/Fe]=0.4. These models were shifted to a
reddening  of $E(B-V)$=0.08, 0.09 and 0.10 and distance of 23.67 kpc calculated in $\S
\ref{sec:redd}$ and $\S \ref{sec:RRLstars}$. Given the scatter in the RGB this
uncertainty in the reddening is not unreasonable.

\begin{figure} 
\includegraphics[width=8.5cm,height=8.5cm]{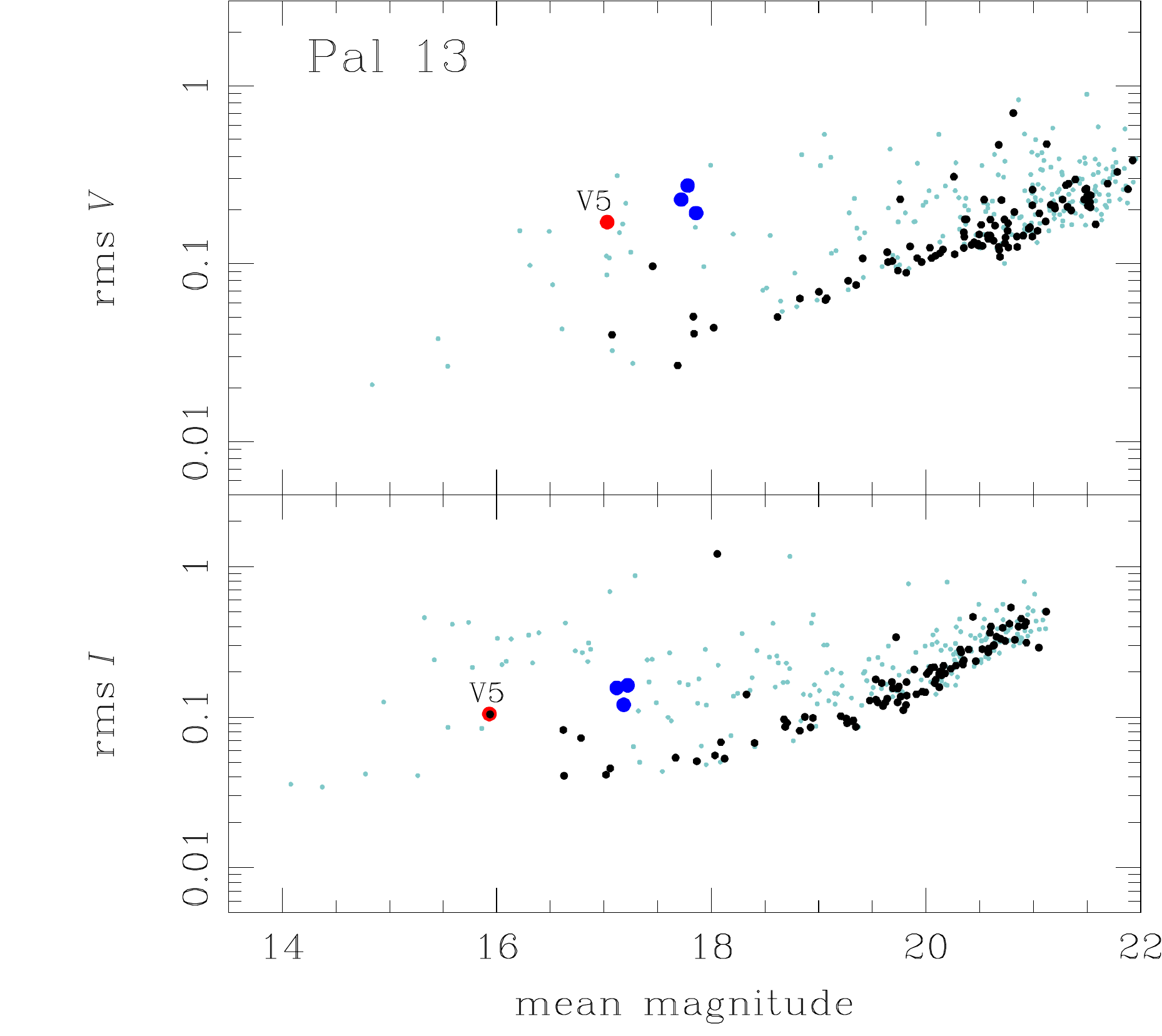}
\caption{$V$ and $I$ rms diagrams. Blue dots represent the known RRab stars V1, V2 and
V3. Black and light blue dots are cluster members and non-members respectively. The
one member star simultaneously outstanding in the $V$ and $I$ bands turned out to be a
RG variable (see its light curve in Fig. \ref{V5}) and we have named it V5.}
    \label{rms}
\end{figure}

\begin{figure} 
\includegraphics[width=8.5cm,height=8.5cm]{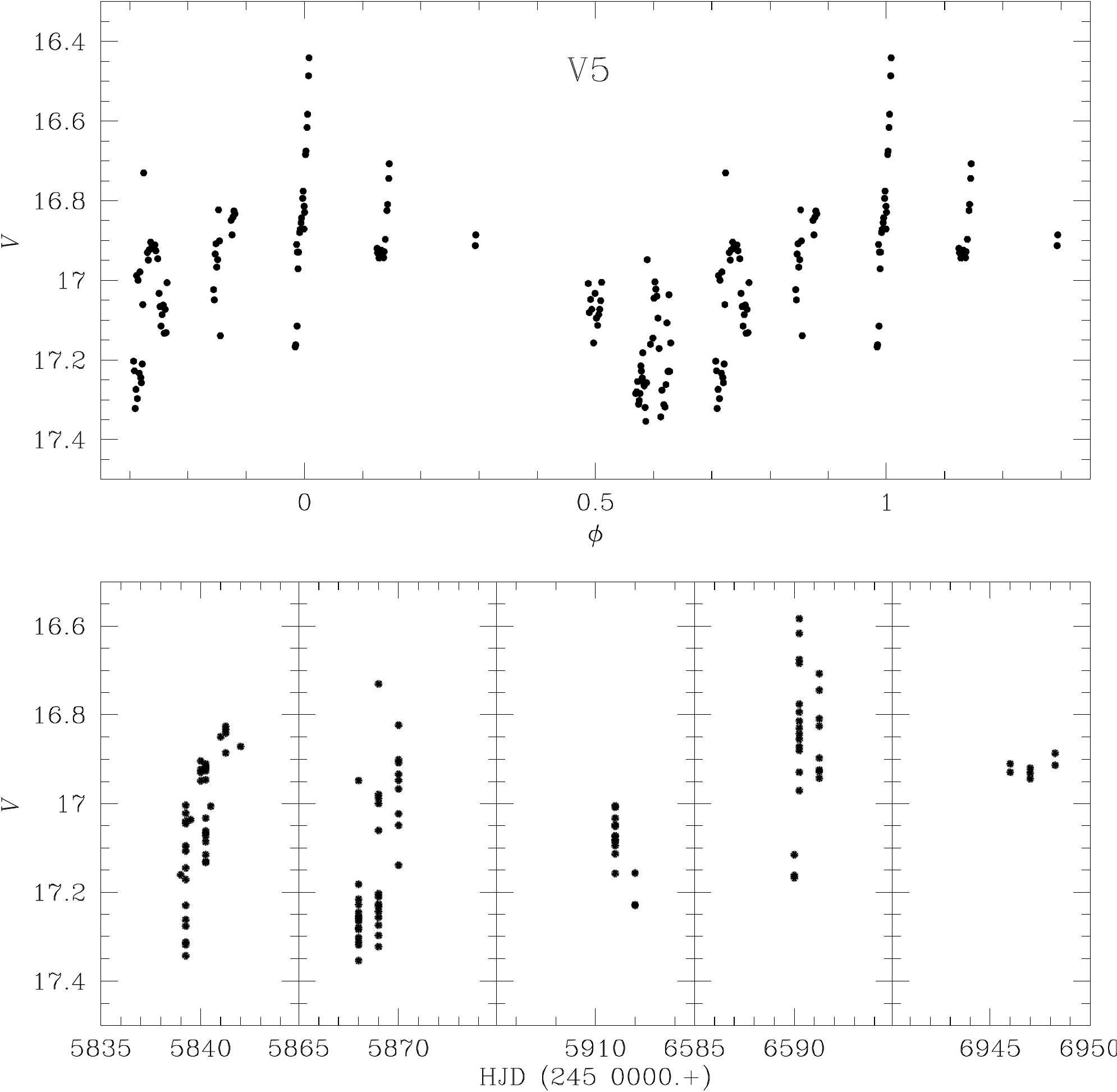}
\caption{V5 $V$-light curve phased with period of 7.26 days. The bottom panels show
the $V$ magnitudes as function of HJD where the variations are very clear.}
    \label{V5}
\end{figure}

\subsection{Previous estimates of the distance of Pal 13}

Literature distance estimates for Pal 13 are generally in good agreement within a
small range. \cite{Borissova1997} reports distance modulus equivalent to distances
between 24.5 and 26.0 kpc. On the other hand, \cite{Siegel2001} and \cite{Cote2002}
find 24.8 and 24.3$\pm$1.3 kpc respectively. The value listed by \cite{Harris1996} is
26 kpc. Our values reported in $\S$ \ref{sec:distance} via independent methods; 23.97
$\pm$ 0.67 and 24.34 $\pm$ 0.50, are in good terms with previous estimates. 

\subsection{Note on the age of Pal 13}

One can find in the literature several estimates of the age of Pal 13. A differential
approach with Pal 5 and an statistical analysis of the CMD led \cite{Borissova1997} to
conclude an age of 12$\pm$2 Gyrs and a metallicity of [Fe/H]=$-1.66$. \cite{Cote2002}
argue a good isochrone fitting with a [Fe/H]=$-1.78$ model of 14 Gyrs on the $V/(B-V)$
CMD, and \cite{Siegel2001} also find that 12 Gyrs is an age consistent with their
$V/(B-V)$ CMD. The calibration of the age indicator $\Delta$V (HB-TO) of
\cite{Peterson1987} (his eq. 3), and our $\Delta$V (HB-TO) = 3.25 estimation from Fig.
\ref{sec:CMD} suggests and age of 12.3 Gyrs. As argued in $\S$ \ref{sec:CMD} our
$V/(V-I)$ CMD also favours an age of about 12 Gyrs. 

\section{The horizontal branch of Pal 13} 

In order to better understand the evolutionary status of the variable stars found on
the 
horizontal branch of Pal 13, we computed our own evolution models, using our well
calibrated 
version of the Eggleton code \citep[see][]{Eggleton1971, Eggleton1972,
Eggleton1973,KPS1997}.
For the mass-loss on the upper RGB, we use the prescription and calibration,
physically
motivated by \cite{KPS2005}.

\begin{figure} 
\includegraphics[width=8.5cm,height=8.5cm]{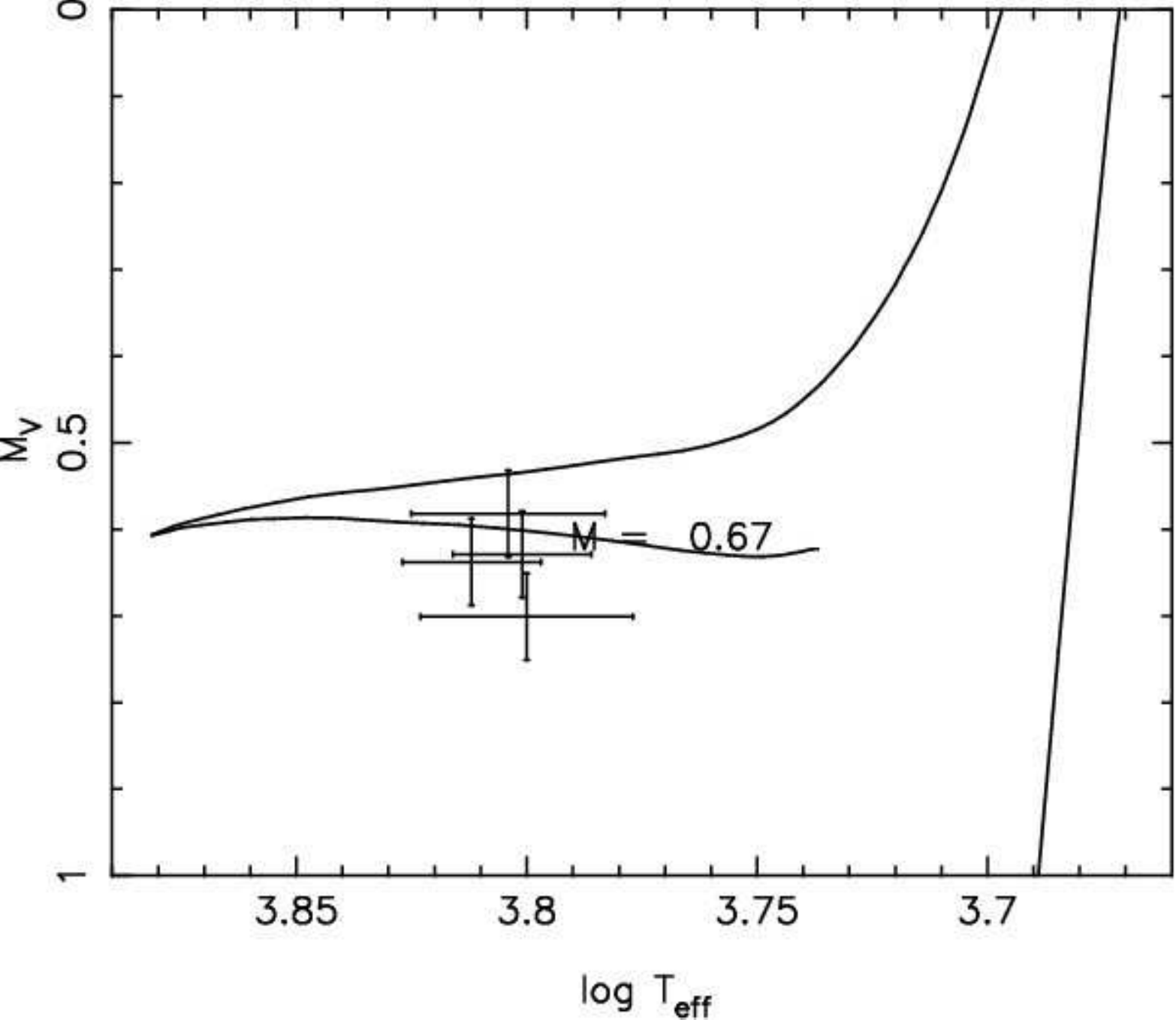}
\caption{Empirically determined physical parameters $M_v$ and $T_{\rm eff}$ of the
observed cluster variables V1 to V4 (see Table \ref{fisicos}) compared with our
evolution track, generated with a ZAHB mass of 0.67 $M_\odot$, consistent
with the average mass obtained from V1 to V4.}
\label{TRACK}
\end{figure}

We find a good match for the stretch of the horizontal branch of Pal 13 in colour (or
rather, effective temperature) by a ZAHB (zero age horizontal branch) model of 0.67
solar masses, of which 0.50 solar masses are concentrated in the He-core, see Fig.
\ref{TRACK}. This is entirely consistent with the average mass we find empirically for
the variable stars V1 to V4 (see Table \ref{fisicos}) and the masses of V1 and V2, in
particular.

Our evolution
models of the respective progenitor stars have an initial zero age main sequence
(ZAMS) mass of 0.85 solar masses. By a 
modified Reimers wind (not dust-driven mass-loss, with additional terms for gravity
and 
effective temperature according to \cite{KPS2005}, this model loses 0.18 solar 
masses on the upper RGB, before it reaches its tip and in reality would undergo a
Helium flash
to lift the degeneracy of its core at an age of 11.95 Gyrs. This age is perfectly
consistent 
with the matching isochrone of VandenBerg (2014) in Fig. 5 and not much different from
the 
other ages found in the literature, as stated above. We consider the He-flash
imminent, and 
therefore ignore the further numerical evolution as unphysical, at a luminosity
provided by 
the onset of He-burning of 0.15 solar luminosities.       

The not very blue horizontal branch of Pal 13 can, hence, be reproduced with mainly
one
evolution model only. The slowest evolution, providing the largest probability of
finding
horizontal branch stars here (and so to explain the nature of V1-V4), is the lower
strech 
in the earlier phases of central helium burning. These stars evolve from just redwards
of 
the instability strip, where they start Helium burning as ZAHB stars, to its blue
side.
Their evolution should be too slow to change their period measurably in human
life-times. 

After central Helium burning, on their return to the red side of the HRD and onto the
AGB, 
HB evolution accelerates. Stars on that upper part of the track would distinguish 
themselves by being brighter by a quarter of a magnitude and a very slow Period
increase, which
might be observable over decades. However, their population number is much reduced by
the 
higher evolutionary speed, and not many of these 
stars should be found. Few significant examples have been identified in the globular
clusters M5 \citep{Arellano2016} and NGC 6171 \citep{Arellano2018_N6171}. Considering
the sparse population of Pal 13, indeed hardly any HB star should be expected to be
found in that phase.

\section{Search for new variables}

Given the clear scarcity of known variables in Pal 13, we have taken advantage of our
time-series photometry to perform a careful search of new variables by different
approaches, notwithstanding the risk of not finding any.

The red box in CMD is an arbitrarily defined blue stragglers region in which 8 cluster
members are identified. An exploration of the light curves of these stars revealed no
variations.

\begin{figure} 
\includegraphics[width=9.0cm]{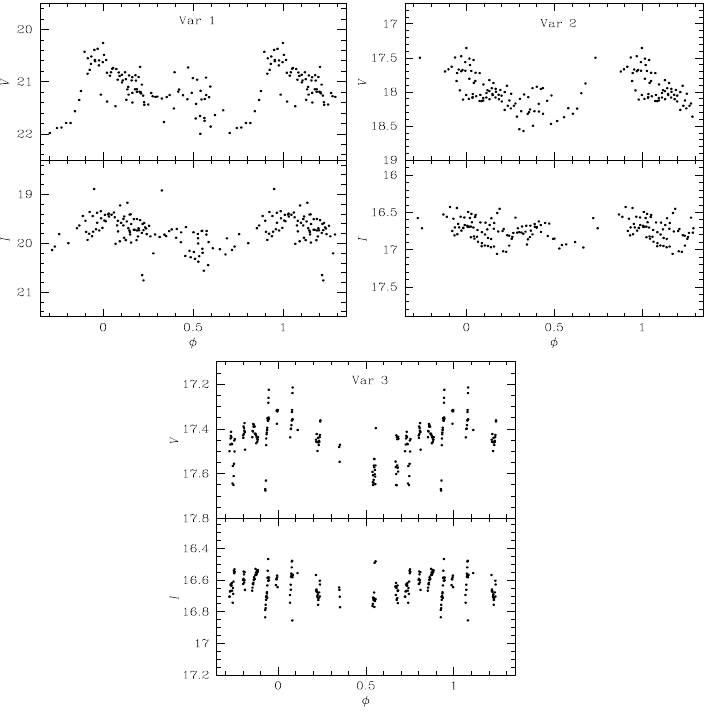}
\caption{Three new variables in the FoV of Pal 13. None is a cluster member. $Var1$
and $Var2$ seem to be RRab stars while $Var3$ is likely a W Virginis star or CW.}
    \label{3vars}
\end{figure}

Fig. \ref{rms} shows the rms magnitude deviation diagrams for all stars measured in
our images, in the field of Pal 13 in the  $V$ and $I$ light curves. Stars with a
large rms are prone to be variable. The one cluster member star that displays a large
rms in both $V$ and $I$ turned our to be a RG variable whose light curve is shown in
Fig. \ref{V5} and we have named it V5.

We have also applied the phase dispersion minimization approach to all the stars in
the FoV of our images: This indeed recovered the four known RR Lyrae stars and
identified three new variables among the non-members in the FoV the cluster. Their $V$
light curves are show in Fig. \ref {3vars} . We refrain from assigning them a variable
number but identify them as $Var1$, $Var2$ and $Var3$. Given their periods and light
curve shapes, $Var1$ and $Var2$ seem to be  RRab stars and Var3 is likely a W Virginis
star or CW. Their ephemerides are given in Table \ref{variables}.

Finally, we blinked all the differential images in our collection in search for
observable flux variations. We identified all variables mentioned above but failed
detecting a new one.

\subsection{Comments on the new variables}

V5. This variable sits on the lower RGB, rather the subgiant branch. Its period of
about 7 days is however too short to classify the star as an SR which are normally
much brighter,  closer to the tip of the RGB, and have periods larger than 20 days.
This kind of variable giants have been found in 47 Tuc \citep{Albrow2001} and $\omega$
Cen \citep{Weldrake2007}. In the Catalogue of Variable Stars in Globular Clusters
these stars are identified as L? (C. Clement. private communication)

$Var1$ and $Var2$. The periods and light curve shape of these two stars suggest they
are RRab stars. Their peculiar position by at least 0.3 mag to the red of the RGB is
puzzling since even being field stars  they would need to be subject to a much heavier
reddening than the rest of the cluster stars.

Var3. The period of 7.5 days and its position on the CMD
suggest this non-member star to be a CW variable, also known as W Virginis stars,
projected on the cluster field.

\section{On the velocity and proper motion dispersion in Pal 13}
\label{evaporation}

Two properties make of Pal 13 a peculiar globular cluster; its shallow surface density
profile and a high velocity dispersion of $2.2 \pm
0.4$ km s$^{-1}$
that results in the M/L of about 40, the highest known among galactic globular
clusters \citep{Siegel2001, Cote2002}. It was shown by \cite{Bradford2011} that
removing binary stars reduces, at least partially, the velocity dispersion, from where
these authors conclude that Pal 13 must have a larger fraction of binaries than most
Milky Way globular clusters. However, the large M/L ratio does not preclude the
possibility that Pal 13 is in fact an ultra-faint dwarf galaxy  or dSph satellite.
From their N-body computations \cite{Kupper2011} concluded that the cluster is near
its apogalacticon, and that the large velocity dispersion is due  to unbound stars
within the cluster  and
extra tidal stars (tidal debries) which get pushed back into the vicinity of the
cluster when the cluster-tail system gets decelerated on its way to apogalacticon, and
not to actual tidal shocking. Such an scenario should be reflected in the proper
motion distribution.

\begin{figure} 
\center{
\includegraphics[width=8.0cm]{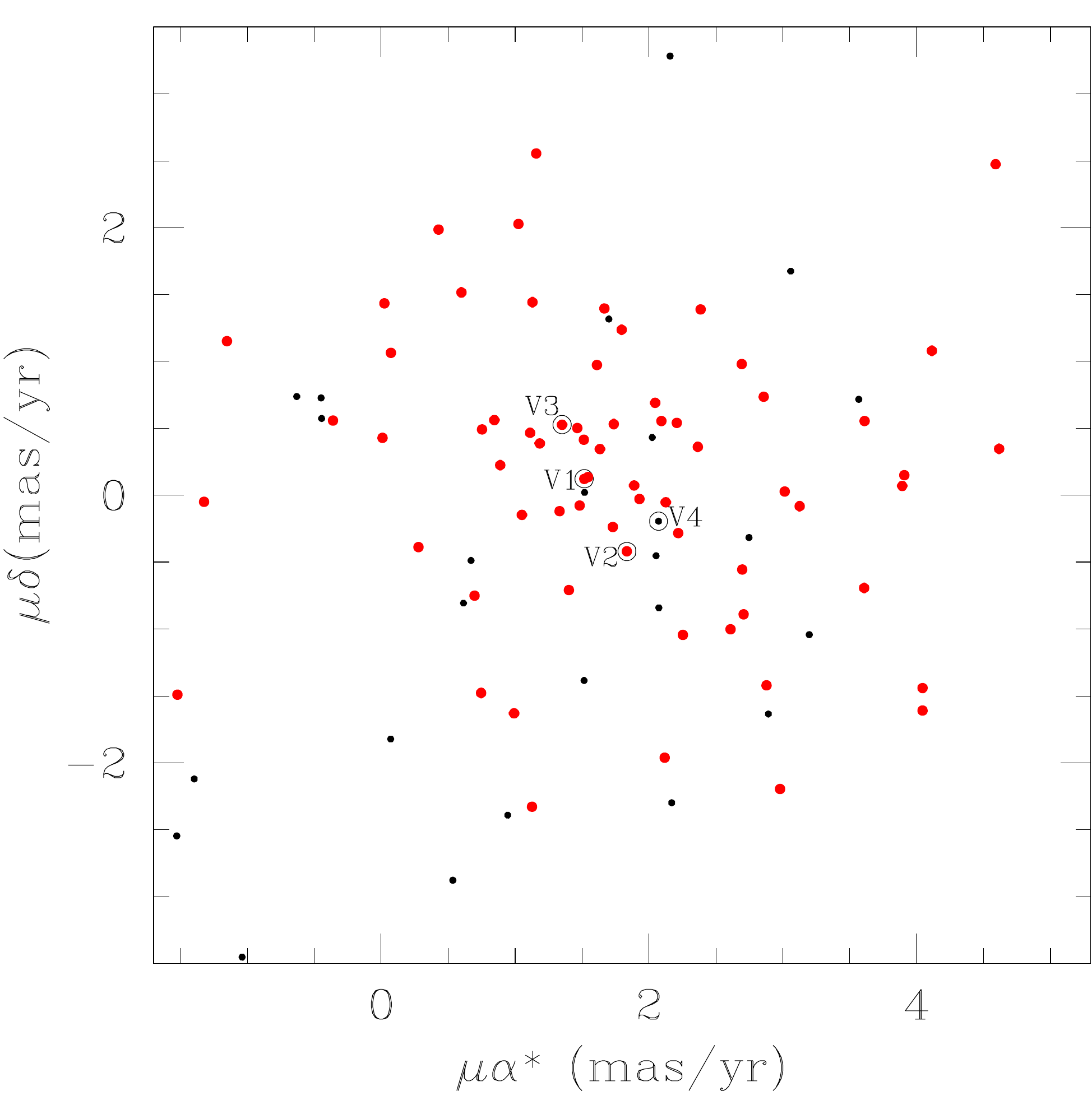}
\caption{Vector Point Diagram (VPD) of Pal 13 based on Gaia-DR2 proper motions. Red
circles represent probable cluster members. The four known RR Lyrae are marked. The
case of V4 is discussed in $\S$ \ref{sec:distance}.}
    \label{Pal13_VPD}}
\end{figure}

We have  extracted from the \cite{Gaia2018}, the proper motions of stars in the field
of Pal 13 within a radius of 8 arc minutes, containing 323 Gaia sources. Bustos Fierro
\& Calder\'on (private communication) found 69 probable members following a method
developed by them and which shall be published elsewhere.
Fig. \ref{Pal13_VPD} shows the vector point diagram (VPD) of Pal 13. The 69 likely
cluster members are represented by the red points. V1-V4 are marked. The average
proper motion components are $\mu \alpha *$= 1.92 $\pm$ 0.19 and $\mu \delta$= 0.13
$\pm$ 0.13. These values can be compared with those of
\citet{Kupper2011} (2.30 $\pm$ 0.26,0.27$\pm$ 0.25) or with more recent values from
the Hubble Space Telescope proper motions calculated by \citet{Sohn2018} (1.70$\pm$
0.09,0.08$\pm$0.06) and those from the $Gaia$-DR2 from \citet{Vasilev2019} (1.615$\pm$
0.101,0.142$\pm$0.089). 

\begin{figure} 
\center{
\includegraphics[width=8.0cm]{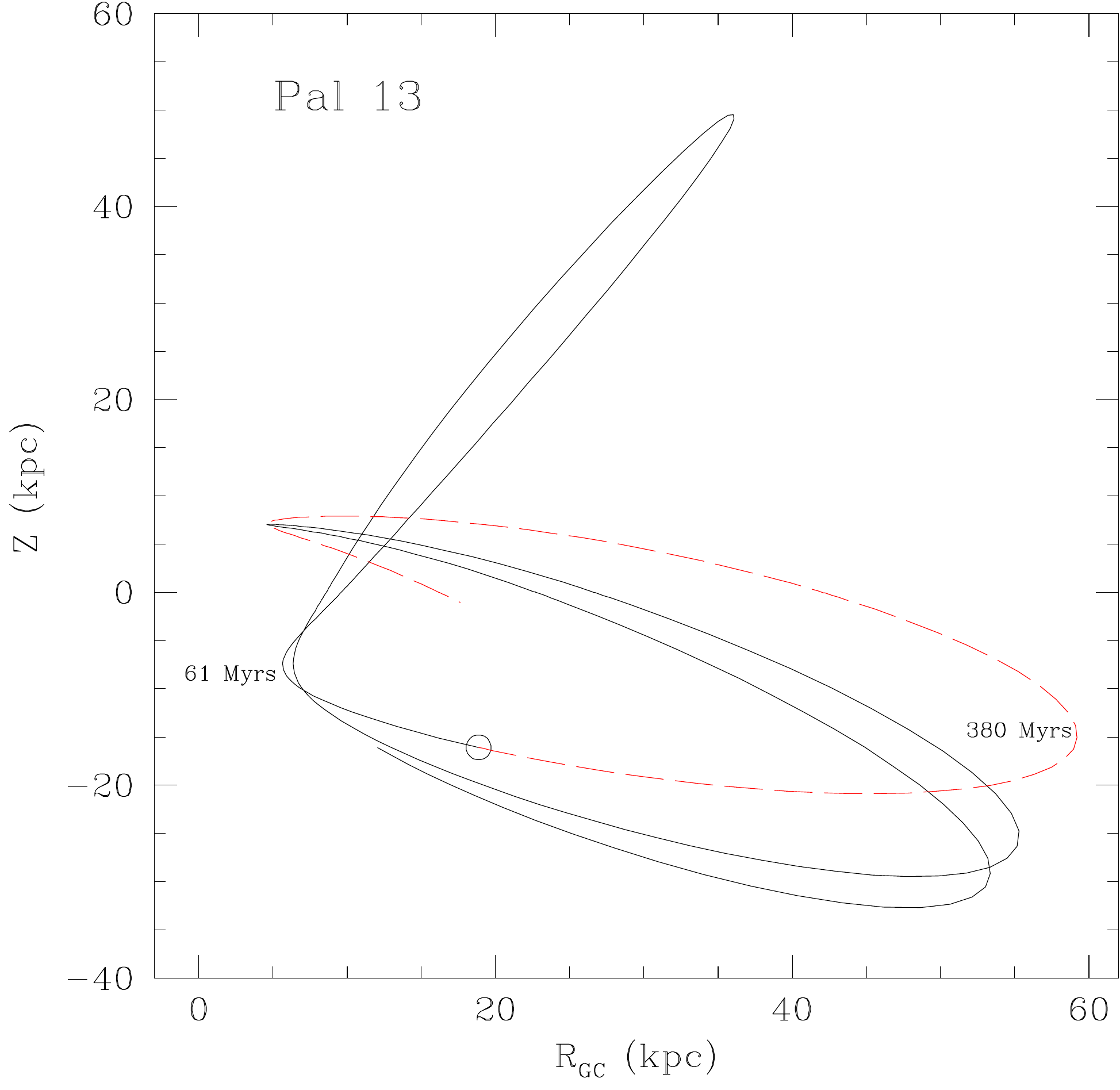}
\caption{Galactic orbit of Pal 13 from 2.5 Gyrs in the past (continuous black curve)
and 0.8 Gyrs into the future (dashed red line). Open circle represents the present
position of the cluster. The timings of the last passage by the perigalacticon and the
next passage by the apogalacticon are indicated relative to the present.}
    \label{ORBITA}}
\end{figure}

By making use of our values for the  mean $Gaia$ proper motions, a mean radial
velocity of 24.5 km/s and a distance to the Sun of 23.67 kpc, in an axisymmetrical
Galactic potential  \citep{Allen2006}, we have calculated the Galactic orbit of Pal 13
from 2.5 Gyrs in the past to 0.8 Gyrs into the future, this is displayed in Fig.
\ref{ORBITA}. For an easy comparison with the orbits in \citep{Kupper2011} (their
figure 2), we used the same galactic and solar motion parameters. The new smaller mean
average proper motions from $Gaia$ produce a more excentric orbit but otherwise they
are quite similar. The cluster experienced its last perigalacticon passage about 61
Myrs and shall reach its next apogalacticon in another 380 Myrs. It is debatable which
of these two contributes most to the proper motion perturbation
of the cluster stars. 

\begin{figure*} 
\center{
\includegraphics[width=12.0cm]{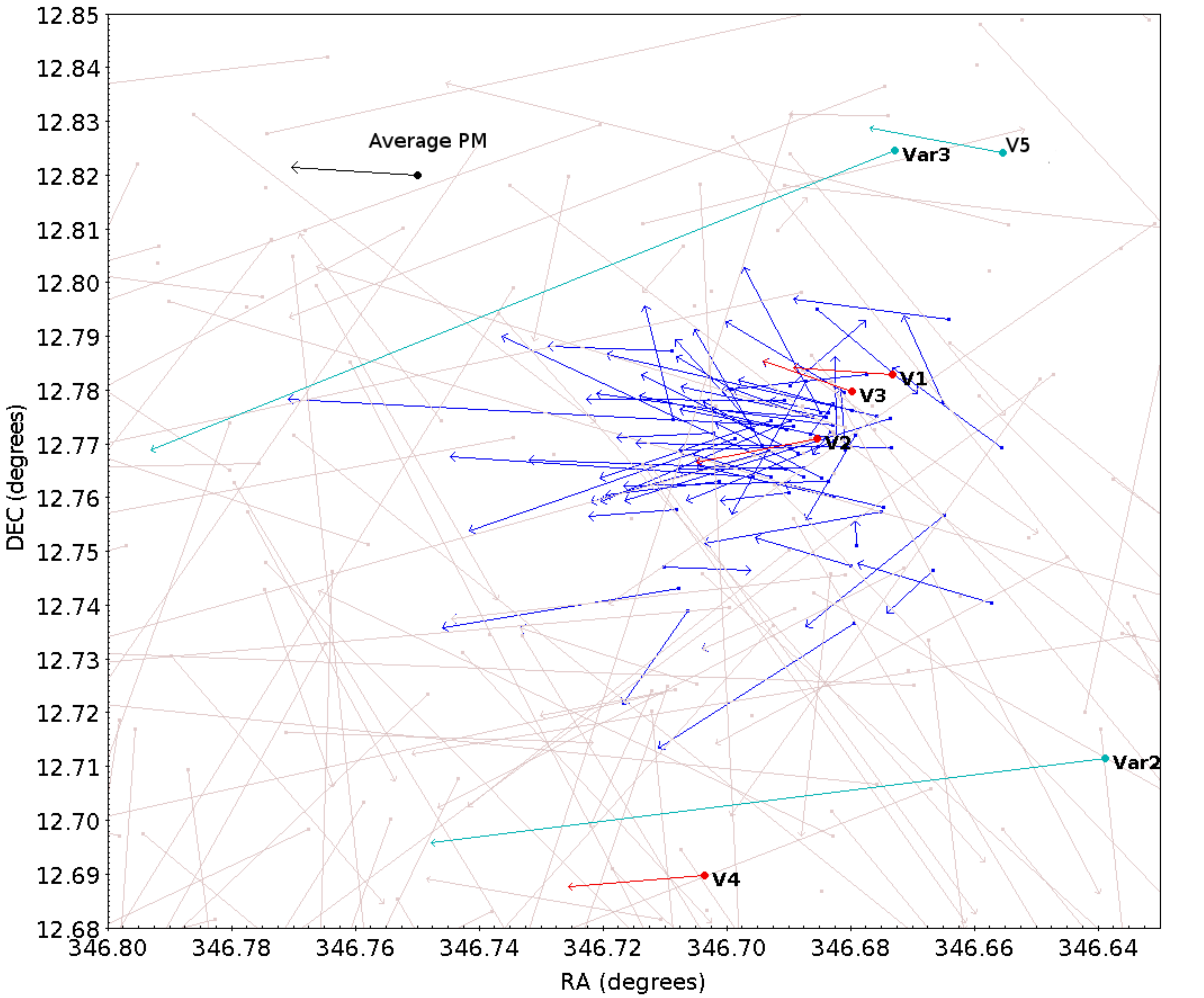}
\caption{Proper motion vectors of stars in the field of Pal 13 (multiplied by an
enlarging factor to make them visible). Blue arrows represent member stars, light gray
non- members, red are the four know RR Lyrae stars and green the newly discovered
variables. The average proper motion of the member stars is indicated by the black
arrow. See $\S$ \ref{evaporation} for a discussion.}
    \label{Pal13_PM}}
\end{figure*}

In Fig. \ref{Pal13_PM}, we have plotted the proper motion vectors on the plane of the
sky. Member and non-member stars are shown with blue and light gray arrows
respectively. The four known RR Lyrae stars are plotted with red arrows and the newly
discovered variables V5, $Var2$ and $Var3$ are shown in green colour. The star $Var1$
has no proper motion in the $Gaia$-DR2 data base.
The average proper motion of the member stars is shown with a black arrow.

A noticeable high dispersion of proper motions even among members is evident. It is
contrary to the dispersion observed in other clusters. To highlight this fact we have
compared with M13, a massive compact cluster whose star members are moving in an
ordered tandem.

\begin{figure*} 
\center{
\includegraphics[width=15.0cm]{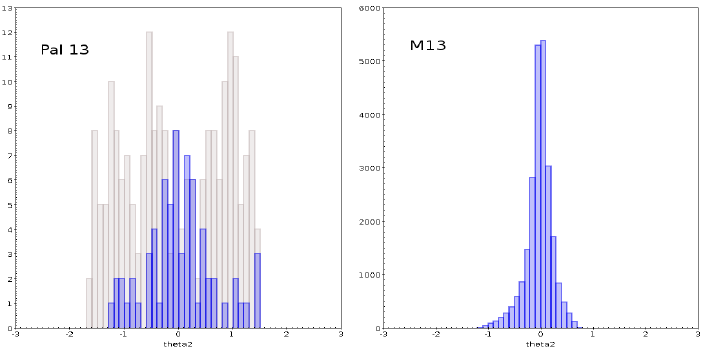}
\caption{Histograms of the proper motion deviations form the average for cluster
members. Left panel illustrates the case of Pal 13 and the right panel for M13. The
larger scatter in the proper motions of Pal 13 is evident. Blue bars represent cluster
members and light gray the non-members. See $\S$ \ref{evaporation} for a discussion.}
    \label{Histograms}}
\end{figure*}

We have calculated the deviation of each proper motion vector relative to the average
of the member stars. Fig \ref{Histograms} shows the distribution of the proper motion
dispersions in Pal 13 and in M13. It seems clear that the proper motions of cluster
members are largely scattered, which is consistent with the scenario of the cluster
being evaporating. While it is true that the cluster is approaching the apogalacticon
as noticed by \cite{Kupper2011}, the cluster is much closer to its last passage by the
perigalacticon, event that has most likely contributed to the proper motions
dispersion, as it has been discussed by \citet{Webb2014a,Webb2014b} in terms of
escaping stars by tidal compression at perigalacticon.

On the basis of the proper motion scatter displayed by Pal 13, and the variable star
positions on the CMD, we can reassure that V1-V5 are cluster members whereas the new
variables $Var2$ and $Var3$ are not.

\section{Summary and discussion}
\label{summary}

The Fourier decomposition of the light curves of the RR Lyrae stars in Pal 13, and the
calibrations and zero points available in recent literature, enable the determination
of the mean metallicity and distance of this faint cluster. [Fe/H]$_{\rm
ZW}=-1.65\pm0.15$ and $23.67\pm 0.57$ kpc were found. Individual values of radius and
mass are also provided. The employment of the $I-band$ RR Lyrae P-L relation derived
by \cite{Catelan2004} gives a distance of 24.34$\pm0.50$ kpc.

\begin{figure*} 
\includegraphics[width=18.0cm]{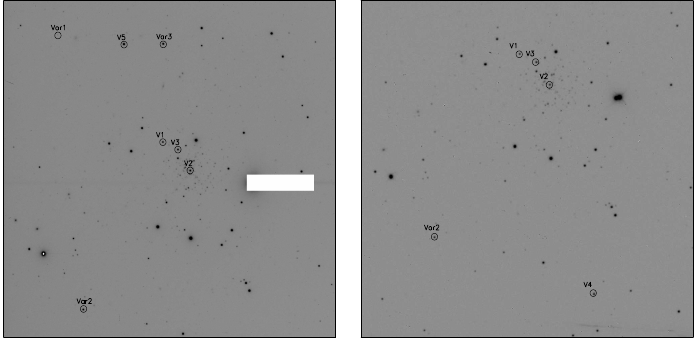}
\caption{Variable stars in the FoV of Pal 13. The left and the right frames
correspond 
to the reference images for Hanle and SPM respectively. North is up and east is to the
right. In the left panel the brightest object was blocked to diminish the "light
bleeding" contamination.}
    \label{chart}
\end{figure*}

In connection with the membership of V4 to Pal 13 we should note that the star is
within 
the King tidal radius \citep{Bradford2011} and its distance matches that of the other
RR Lyrae in the cluster. Furthermore, an analysis of the proper motions of Pal 13 and
surrounding regions, in particular of those considered members of the cluster, shows
that V4's proper motion is within the members scatter, consistent with being part of
cluster. 

We note that in the HB instability strip of Pal 13, the RRab stars V1, V2, V3 and V4 
lie on the fundamental region of the instability strip. They are in the central
He-burning phase and very slowly evolving bluewards on the HB. Pal 13 is yet another
Oo I cluster without fundamental pulsators in the "either-or" region, property that it
is shared with most, but not all, Oo I clusters, and certainly with all OoII clusters
(see \cite{Arellano2018_N6362} for a recent discussion on this issue).

We report the discovery of a new variable, V5 sitting on the RGB with a period of about 7.2 d, classified as a bright giant variable or probably an SR. Three more variables were found among the non-member population of the cluster; we have refrained from assigning a formal variable number to them but identified as $Var1$, $Var2$ and $Var3$, in the finding chart of Fig. \ref{chart} and in Table \ref{variables}, and classified the first two as RRab and as CW respectively.

Finally, the inspection of the GAIA proper motions of stars considered members, 
reveals a large scatter, confirming that the cluster members are dissociating, in
accordance with the scenario that the internal dynamics is altered as the cluster is
near its apogalacticon \citep{Kupper2011}. 

\section*{Acknowledgments}
We are thankful to Dan Deras for his help during some of the observations, to Ivan
Bustos 
and Jes\'us Calder\'on for their useful comments on the stellar membership and to
Alonso Luna for his help with the production of some graphical material. We
acknowledge the financial support from DGAPA-UNAM (Mexico) via grant IN106615-17. M.
A. Yepez is grateful to the Department of Astronomy of the University of Guanajuato
for warm hospitality during the preparation of this work. We have made an extensive
use of the SIMBAD and ADS services, for which we are thankful.

\bibliography{bibliografia}

\end{document}